# A Broad-Spectrum Diffractive Network via Ensemble Learning

JIASHUO SHI,[1,2,4] YINGSHI CHEN,[3,4] AND XINYU ZHANG[1,2,*]

[1]National Key Laboratory of Science and Technology on Multispectral Information Processing, Huazhong University of Science and Technology, Wuhan 430074, China
[2]School of Artificial Intelligence and Automation, Huazhong University of Science & Technology, Wuhan 430074, China
[3]Giga Design Automation Co., Ltd. 1801 Shahe West Road, Nanshan District, Shenzhen 518055, China
[4]Co-first authors with equal contribution
*Corresponding author: x_yzhang@hust.edu.cn

**We proposed a broad-spectrum diffractive deep neural network (BS-D$^2$NN) framework, which incorporates multi-wavelength channels of input lightfields and performs a parallel phase-only modulation utilizing a layered passive mask architecture. A complementary multi-channel base learner cluster is formed in a homogeneous ensemble framework based on the diffractive dispersion during lightwave modulation. In addition, both the optical Sum operation and the Hybrid (optical-electronic) Maxout operation are performed for motivating the BS-D$^2$NN to learn and construct a mapping between input lightfields and truth labels under heterochromatic ambient lighting. The BS-D$^2$NN can be trained using deep learning algorithms so as to perform a kind of wavelength-insensitive high-accuracy object classification.**

By transplanting the operating rules of the conventional electronic-based neural network, the optical-based neural network has been reported through different implementations [1-3] with several obvious advantages in transmission speed, power consumption, and parallelization capability. As one of the representatives in the field of deep learning with optics, the diffractive deep neural network (D$^2$NN), which is a kind of materialized multilayer diffractive mask, has received more attention in recent years [4-8]. Utilizing a diffraction model with a backpropagation and gradient descent algorithm, the D$^2$NN can be trained using dozens of training data in an electronic computer. Then the optimized network parameters can be introduced into a layered architecture formed by 3D-printing, two-photon direct laser writing, or conventional wet etching [9,10]. The effectiveness in dealing with large-scale information processing tasks, such as image classification, pulse shaping, and saliency detection [11-13], has been verified by constructing a suitable optical neural network.

As demonstrated, almost all D$^2$NNs are already designed according to incident lightwave signals with a single wavelength due to an obvious prediction deterioration caused by the diffraction dispersion of the different wavelength components of incident beams in network forward propagation. Especially, the phase plate, as a vital component of the D$^2$NN, presents a dramatical wavelength distinguishing, which can be further enhanced through remarkably expanding the wavelength difference or range of incident beams and then the number of network layers. Deriving the fundamental principle based on conventional achromatic lens, this Letter proposed a broad-spectrum diffractive deep neural network (BS-D$^2$NN) framework, and then demonstrated a robust inference with a heterochromatic incidence in performing image classification task. Interestingly, we found that several discrete points of the incident lightfield in frequency domain will remarkably prompt the D$^2$NN to construct a multiple-base learner jointly optimized mechanism similar to Random Forest strategy, where each base learner independently processes the associated monochromatic lightfields of targets and then outputs their predictions. All opinions are then synthesized by the decision-making layer, and thus the final prediction of the whole system is given efficiently. As stated, the network mapping differences originated from the beam dispersion are enough to construct an ensemble learning configuration of the BS-D$^2$NN, in which each base learner maintains a diversity prediction attributing to improve the system inference performance. Relying on the above mechanism, the BS-D$^2$NN has numerically achieved a test accuracy of more than 90.08% over Fashion-MNIST dataset under a heterochromatic illumination.

As a multi-channel signal ensemble processing architecture, the BS-D$^2$NN needs to employ a parallelism combination paradigm such as the Bootstrap Aggregating strategy [14,15] analogous to conventional electronic machine learning to aggregate the inference results of the base learners. It should be noted that the multi-channel attributing to the BS-D$^2$NN refers to multiple discrete lightwave frequencies, but the typical RGB color channels in conventional electronic network models having not any ensemble learning configuration because of no network wavelength or frequency difference recognition and manipulation function. Considering the limitations of all-optical computing, an optical version of the Sum and the Hybrid Maxout Operations for realizing a hierarchical computation with a featured high-throughput and ultra-low energy consumption is also presented. The Sum Operation layer is used to significantly extend the wavelength difference or range processed by the BS-D²NN, and consequently improve the accuracy of ~1.04% compared to that of the plain D²NN, which only works in an expert wavelength by default. On the other hand, the operation of employing the Hybrid Maxout Operation layer in performing network training and prediction can further improve the test accuracy of ~5.74% compared to that of the plain D²NN even though no effective improvement in robustness. Finally, this study shows a novel method of achieving base learner ensemble strategy based on an obvious chromatic aberration in lightfield propagation in a layered diffractive network. We believe that the proposed framework might fuel the further development of machine learning in multi-spectral information processing applications.

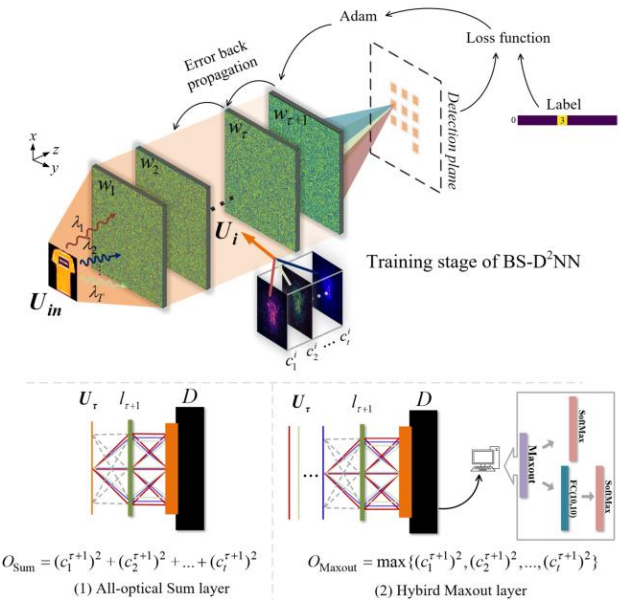

Fig. 1. Training process and operation principle of the BS-D²NN. Taking advantage of the key manipulations including: error backpropagation, stochastic gradient descent, and activation function, the output lightfields are used to optimize the neuron weights of the BS-D²NN.

To perform the broad-spectrum training, i.e., the multi-frequency channel jointly optimizing and aggregating of the base learners with a distinct monochromatic illumination, the operation principle of the BS-D²NN is illustrated in Fig. 1. As shown in the figure, $U_{in}$ represents the lightfield incident upon the input aperture with a short sleeve pattern selected as a typical target, and then $U_i$ is a mixing of multiple monochromatic lightfields outfrom the $i^{th}$ layer. The incident wavelengths are defined as a set of $\{\lambda_t \in R^+ \mid t = 1,2,...,T\}$, where $T$ denotes the number of the all participating frequency points. In detail, $U_i$ impinges on the $(i+1)^{th}$ network layer after going through free space diffraction, where the median $\lambda_i$ in the wavelengths set above is selected as an indicator, and then the layer spacing $25\lambda_i$. The Rayleigh-Sommerfeld diffraction operator $H_i$ is utilized to describe the lightfield propagation with a diffraction distance $z_i$ along z-axial direction, and the wave vector $\vec{k}_t = 2\pi\vec{n}/\lambda_t$, and the pixel distance $r_i$. Both $H_i$ and $U_i$ can be written as

$$H_i = F[\frac{z_i \exp(jk_t r_i)}{j\lambda_t r_i^2}], \qquad (1)$$

$$\hat{U}_i = F^{-1}[F[U_i]H_i], \qquad (2)$$

where $F[.]$ and $F^{-1}[.]$ represent the Fourier transformation and inverse Fourier transformation, respectively, and $\hat{U}_i$ is the diffractive lightfields formed.

After the incident lightfields propagating through the $i^{th}$ layer of the BS-D²NN, all light frequency channels will be simultaneously modulated with distinct phase modulation coefficients $w_i$, which satisfies the relations of

$$w_i = \frac{2\pi(n(\lambda_t) - n_{air})\Delta v_i}{\lambda_t}, \qquad (3)$$

$$U_{i+1} = \hat{U}_i \exp(jw_{i+1}), \qquad (4)$$

where $n(\lambda_t)$ denotes the refractive index of the structural material of the BS-D²NN, and $n_{air}$ the refractive index of air, and $\Delta v_i$ the relative height map of the $i^{th}$ network layer. Then the multi-frequency field $U_{i+1}$, as a combination of all monochromatic lightfields and having different amplitude and phase distributions due to different phase modulation coefficients for associated illumination wavelength of sub-lightfields, is constructed. So, the lightfields reaching the activation layer, will be processed by the Sum Operation or the Hybrid Maxout Operation demonstrated by the lower subplot in Fig. 1, after continuous multilayer phase modulation. The diffractive network outputs from the activation layers are given by

$$O_{Sum} = (c_1^{\tau+1})^2 + (c_2^{\tau+1})^2 + ... + (c_t^{\tau+1})^2, \qquad (5)$$

$$O_{Maxout} = \max\{(c_1^{\tau+1})^2,(c_2^{\tau+1})^2,...,(c_t^{\tau+1})^2\}, \qquad (6)$$

where $O_{Sum}$ and $O_{Maxout}$ denote the output lightfields of the BS-D²NN with the Sum Operation and the Hybrid Maxout Operation, respectively. $c_i^{\tau+1}$ represents the output of the $i^{th}$ frequency channel of the $(\tau+1)^{th}$ layer. During the training stage corresponding to an electronic computer, both $O_{Sum}$ and $O_{Maxout}$ from the activation layer are the indicators for performing target recognition. The light intensity of a specific ten areas presenting the probability of ten categories is converted into one-hot sequences, and then enters the loss function together with the ground truth label for calculating the prediction error during performing back propagation and stochastic gradient descent. After multi-batches of training, the network learns an approximate mapping relationship from input lightfields to labels, so as to implement a high-accuracy broad-spectrum prediction by configuring the optimized diffractive layers.

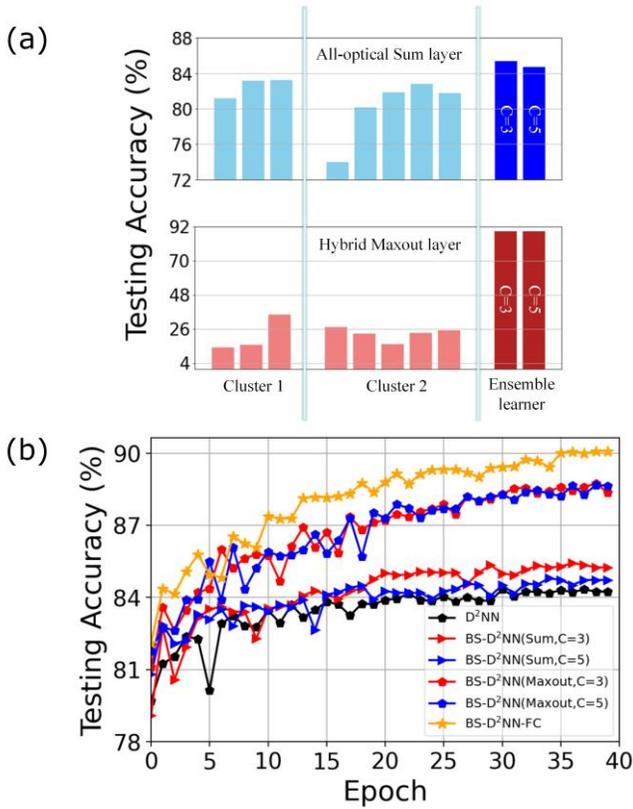

Fig. 2. Performance comparison based on the prediction accuracy for associated network models. (a) Aggregating strategy effect for ensemble learners. (b) Classification task prediction comparison of both the plain D²NN and the BS-D²NN on Fashion-MNIST test dataset.

As shown, the common Maxout Operation can be easily implemented in traditional electronic computers or analog circuits with almost negligible computational consumption; it is actually a universal approximator for increasing the expressivity of the conventional electronic networks. As proved by [16,17], any continuous function can be approximated arbitrarily well by a Maxout network. On the other hand, the Maxout Operation is especially suitable for the multi-frequency channel aggregating of the diffractive network, which greatly improves the ensemble model's fitting capability. More attractively, the Hybrid Maxout Operation can be realized all-optically by utilizing optical nonlinearity of photorefractive crystal, which can be used to perform parallel optical thresholding comparing and the maximum operations [18,19].

To sum up, the heterochromatic lightfield with the same complex amplitude distribution transmission in a diffractive optical network will bring about different transformations for each monochrome lightfield owing to the dispersion effect of diffractive phase modulation. In other words, all single-frequency incident sub-lightfields can be treated with a distinct featured mapping even though propagating through the same fixed diffractive network, which is similar to the base learner in an ensemble learning strategy. Furthermore, all modulated sub-lightfields will be received by the CMOS/CCD imaging sensors at the same time. Only one monochromatic lightfield can be recorded if other frequency channels trained in diffractive network are unrecognized by the sensors above, which means a conventional D²NN prediction process. For a broad-spectrum detector, all output sub-lightfields will be superimposed in the photosensitive plane, which is equivalent to implement an optical Sum activation after layered diffractive modulation. This multi-frequency aggregating method will further improve the prediction accuracy of the diffractive network so as to adapt to every trained single-frequency incident sub-lightfields. On the other hand, a Hybrid Maxout Operation based on the electronic computers or analog arithmetic circuits and then used in network training and prediction stage will scarify the robustness of the diffractive network to single-frequency lightfields and then computing speed to obtain a higher blind test accuracy.

In detail, the performance comparison of the BS-D²NN and the plain D²NN is demonstrated in Fig. 2. All diffractive networks are trained with 55,000 images (5000 validation images) from the Fashion-MNIST dataset, and are numerically tested using 10,000 images from the Fashion-MNIST test dataset. To evaluate the improvement of the network prediction performance for frequency channel aggregating strategies, the blind test accuracy of the base learner clusters and ensemble learners (i.e. the BS-D²NN with distinct aggregating strategies) are shown in subplot (a), where C denotes the number of the frequency channel in a training process. For example, the case of C=3 represents that the diffractive network receives simultaneously the incident wavelength of 1.5 μm and 1.8 μm and 2.2 μm in training and testing stage. The case of C=5 represents that of incident wavelength of 1.3 μm, 1.5 μm, 1.8 μm, 2.0 μm, and 2.2 μm, simultaneously. Furthermore, each frequency channel can be considered as inducing the network to generate a new base learner. Then the case of C=3 means there are three base learners in a diffractive system, which is represented by Cluster 1. In the same way, Cluster 2 represents five base learners in a BS-D²NN.

When the BS-D²NN employs the Sum aggregating strategy so as to easily perform optical operation for broad-spectrum system, most base learners show an excellent estimate capability in object recognition. The ensemble learner (C=3) achieves a 2.17% test accuracy improvement compared with the average value of the base learner prediction accuracy. As a strong nonlinear operation, the manipulation of utilizing the Maxout layer brings a greater fitting capability and a wider hypothesis space to network, which has been verified numerically to achieve a significant improvement for a BS-D²NN in subplot (a). Compared with the Sum layer, the Hybrid Maxout Operation prompts each base learner to become an under-fitting classifier (e.g., the average test accuracy of Cluster 1 being only 21.66%), which can learn more blurred and more global featured mapping for target lightfields. In addition, we present more model comparisons in subplot (b). In terms of conventional single-frequency $D^2NN$, a 82.03% numerical blind test accuracy under 2.2 μm incident lightfields can be obtained. The prediction performance on the classification task is also improved when the Sum Operation is employed in the last layer of the BS-$D^2NN$, which works in a broad-spectrum mode and aggregates more frequency channel information for estimating the target category. Concretely, the blind testing accuracy of the BS-$D^2NN$ with a Sum Operation for C=3 or C=5 is improved to 85.38% or 84.73%. By comparison, the operation of employing a Maxout activation function in the last layer will further improve the network fitting capability in multi-frequency channels, and a 88.92% or 88.96% blind testing accuracy for two kinds of multi-channel BS-$D^2NN$ can be acquired, respectively. The accuracy is increased to 90.08%, when a fully connected layer is used after the Maxout layer (i.e., the BS-$D^2NN$-FC). Apparently, the under-fitting base learners can better obtain the overall feature of incident lightfields in different frequency channels, which considerably improves the entire network system with the aggregating of the Hybrid Maxout Operation.

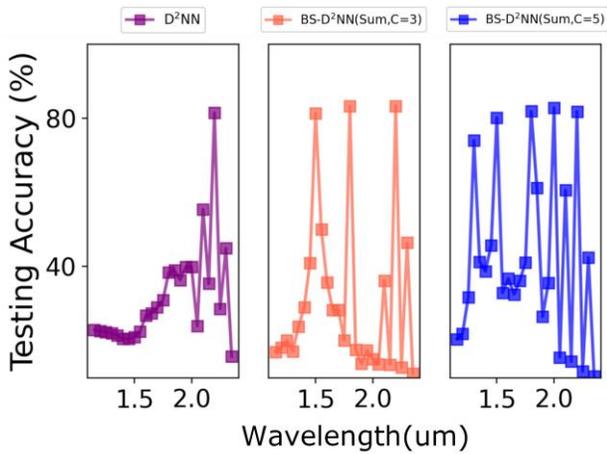

Fig. 3. Wavelength insensitive characteristics of an all-optical BS-$D^2NN$ with a Sum activation function.

Besides, it is empirically verified that the BS-$D^2NN$ with a Sum layer demonstrates an obvious wavelength insensitive characteristics compared with the plain $D^2NN$. We have selected 28 uniform discrete frequency points in a wavelength range from near-infrared to mid-infrared (1.1 μm-2.5 μm) as incident wavelength of the system. The trained model with the same network architecture and optimized parameters shown in Fig. 2(b) can implement a targeted deterministic task. A statistical inference performance is shown in three subplots of Fig. 3. Typically, the prediction accuracy will be gradually decreased in off-center wavelength components due to the modulation coefficient varying of the phase mask to the incident lightfields. The inference capability of the plain $D^2NN$ will reach its peak at the central wavelength of 2.2 μm. As the wavelength shifting, the accuracy drops sharply and thus tends to a stabilized value of 22.53%, which is higher than that of random prediction. In terms of the BS-$D^2NN$ with a Sum layer, both the peak accuracy and the number of relative peak points are taken into account, especially for the BS-$D^2NN$ (C=5), which has achieved five discrete frequency points with > 74% blind testing accuracy and 42.36% average accuracy in the broad-spectrum mode compared with 31.54% average accuracy of the conventional single frequency $D^2NN$.

In summary, we have introduced a broad-spectrum diffractive network (BS-$D^2NN$), which has combined the fundamental principle of ensemble learning and conventional achromatic diffractive lens. The multi-frequency channels of the incident lightfields promote diffractive network formed by a homogeneous ensemble framework due to the dispersion effect during diffractive phase modulation process. Furthermore, two aggregating strategies are presented and employed in diffractive network to improve the prediction performance of the entire ensemble optical system, which also presents a wavelength insensitive property to highlight the development towards a kind of multi-spectral intelligent optical device.

**Funding**

**Disclosures**

The authors declare no conflicts of interest.